\documentclass[twocolumn,english,aps,prl,superscriptaddress,address,showpacs,showacknowledgments]{revtex4}
\pdfoutput=1
\usepackage[latin9]{inputenc}
\usepackage{babel}
\usepackage{amsmath}
\usepackage{amssymb}
\usepackage{graphicx}
\usepackage{esint}
\usepackage[unicode=true,
 bookmarks=true,bookmarksnumbered=false,bookmarksopen=false,
 breaklinks=false,pdfborder={0 0 1},backref=false,colorlinks=false]
 {hyperref}
\hypersetup{pdftitle={Pattern phase diagram for 2D arrays of coupled limit-cycle oscillators},
 pdfauthor={Roland Lauter, Christian Brendel, Steven J. M. Habraken, and Florian Marquardt}}

\makeatletter
\@ifundefined{textcolor}{}
{%
 \definecolor{BLACK}{gray}{0}
 \definecolor{WHITE}{gray}{1}
 \definecolor{RED}{rgb}{1,0,0}
 \definecolor{GREEN}{rgb}{0,1,0}
 \definecolor{BLUE}{rgb}{0,0,1}
 \definecolor{CYAN}{cmyk}{1,0,0,0}
 \definecolor{MAGENTA}{cmyk}{0,1,0,0}
 \definecolor{YELLOW}{cmyk}{0,0,1,0}
}

\usepackage{babel}
\usepackage{babel}
\usepackage{babel}
\usepackage{babel}
\usepackage{bbold}
\hypersetup{
   colorlinks=true,       
   linkcolor=red,          
   citecolor=green,        
   filecolor=magenta,      
   urlcolor=blue           
}

\makeatother

\begin{document}

\title{Pattern phase diagram for 2D arrays of coupled limit-cycle oscillators}

\author{Roland Lauter}

\email{Roland.Lauter@physik.uni-erlangen.de}

\selectlanguage{english}%

\affiliation{Institut f\"ur Theoretische Physik II, Friedrich-Alexander-Universit\"at Erlangen-N\"urnberg, Staudtstr. 7, 91058 Erlangen, Germany}

\affiliation{Max Planck Institute for the Science of Light, G\"unther-Scharowsky-Stra{\ss}e
1/Bau 24, 91058 Erlangen, Germany}

\author{Christian Brendel}

\affiliation{Institut f\"ur Theoretische Physik II, Friedrich-Alexander-Universit\"at Erlangen-N\"urnberg, Staudtstr. 7, 91058 Erlangen, Germany}

\author{Steven J. M. Habraken}

\affiliation{Institut f\"ur Theoretische Physik II, Friedrich-Alexander-Universit\"at Erlangen-N\"urnberg, Staudtstr. 7, 91058 Erlangen, Germany}

\author{Florian Marquardt}

\affiliation{Institut f\"ur Theoretische Physik II, Friedrich-Alexander-Universit\"at Erlangen-N\"urnberg, Staudtstr. 7, 91058 Erlangen, Germany}

\affiliation{Max Planck Institute for the Science of Light, G\"unther-Scharowsky-Stra{\ss}e
1/Bau 24, 91058 Erlangen, Germany}
\begin{abstract}
Arrays of coupled limit-cycle oscillators represent a paradigmatic
example for studying synchronization and pattern formation. They are
also of direct relevance in the context of currently emerging experiments
on nano- and optomechanical oscillator arrays. We find that the full
dynamical equations for the phase dynamics of such an array go beyond
previously studied Kuramoto-type equations. We analyze the evolution
of the phase field in a two-dimensional array and obtain a ``phase
diagram'' for the resulting stationary and non-stationary patterns.
The possible observation in optomechanical arrays is discussed briefly.
\end{abstract}
\maketitle
Synchronization is an important concept in many branches of physics,
chemistry, biology and other sciences \cite{Synchronization_universal_concept}.
Within the past two years, a number of experiments have demonstrated
for the first time synchronization between two nanomechanical oscillators
\cite{Tang_Sync_racetrack,Roukes_Sync_nanomechanical_oscillators,Lipson_Sync_using_Light}.
These systems are driven through a Hopf bifurcation into a limit-cycle
oscillation, where the energy pump is supplied through feedback or
an optical drive. Future arrays of synchronized mechanical Hopf oscillators,
with more than just two components, promise to provide robustness
against both disorder and noise. Considerable theoretical attention
has recently been devoted to the problem of synchronization in arrays,
both on the general level and for predicting the behavior of specific
systems (e.g. in nanomechanics \cite{Cross_Lifshitz_Sync_frequency_pulling,Cross_Lifshitz_Sync_frequency_pulling_PRE,Milburn_Sync_common_cavity,Heinrich_Marquardt_Collective_Dynamics,Allen_Cross_Frequency_precision_with_spirals,Cross_Improving_Frequency_Precision}
or trapped ion systems \cite{Lee_Cross_Ion_Trapping}). Some progress
has also been made in the quantum regime \cite{Giorgi_Zambrini_Quantum_correlations,Ludwig_Marquardt_Quantum_Many_Body,Mari_Fazio_Measures_of_qu_sync,Walter_Bruder_Quantum_sync,Lee_Sadeghpour_Quantum_sync}.
It is efficient to focus on the dynamics of the crucial phase degree
of freedom, where the most prominent phenomenological model is the
one introduced by Kuramoto \cite{Kuramoto_original,Acebron_Spigler_Review},
which more recently has been supplemented by so-called reactive terms
\cite{Cross_Lifshitz_Sync_frequency_pulling_PRE}. 

In the present work, we will explore synchronization and deterministic
pattern formation \cite{Cross_Pattern_Formation_review} for a two-dimensional
array of identical Hopf oscillators. We present the complete effective
model for the phase dynamics. Starting from the widely applicable
model of coupled limit-cycle oscillators, we find that the classical
phase evolution is affected by extra contributions beyond those investigated
previously. These can have a significant impact on the dynamics. Our
simulations of the effective model reveal various stationary and non-stationary
patterns in different parameter regimes. Phase correlators, length
scales, and macroscopic pattern dynamics will be discussed. These
are relevant for determining whether an array can easily settle into
a phase-locked state, which is important for applications.

\begin{figure}
\includegraphics[width=1\columnwidth]{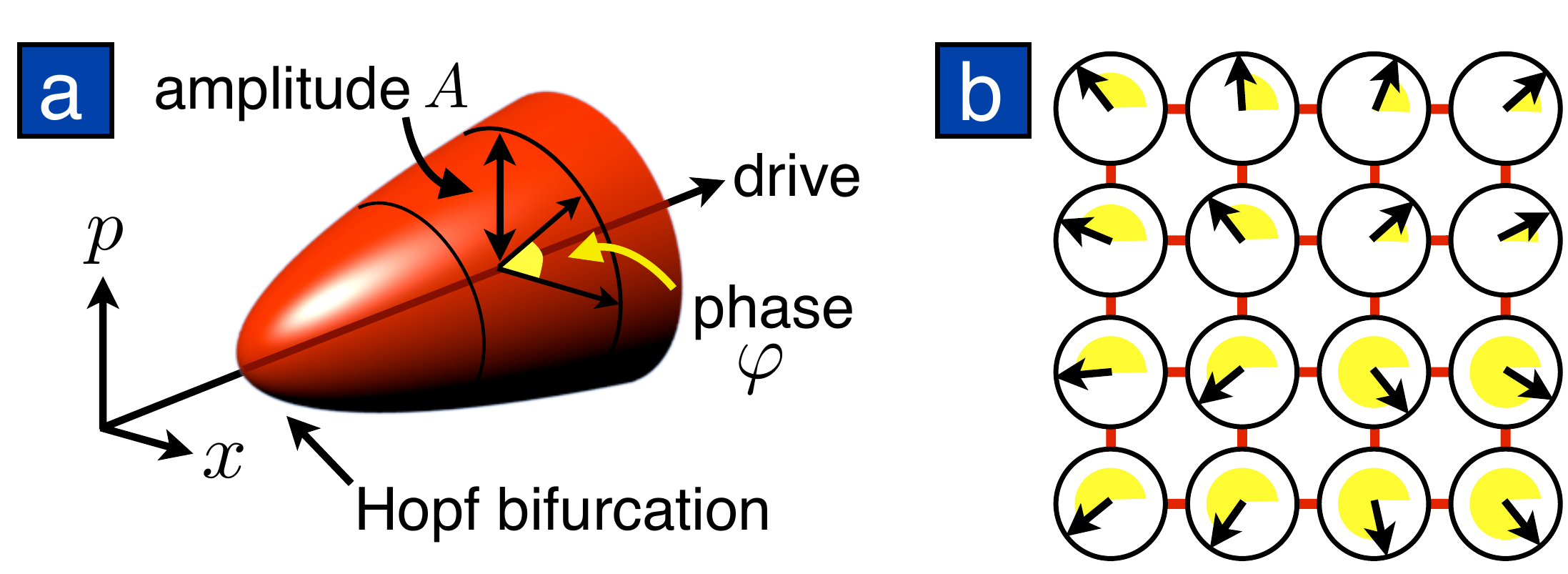}\caption{(a) A single Hopf oscillator undergoes dynamics on a limit cycle,
with an amplitude set by external parameters (such as the drive providing
the power) and a time-evolving phase $\varphi(t)$. (b) Array of coupled
Hopf oscillators. Often, the system can be well described with a single
degree of freedom per lattice site, the phase $\varphi_{i}$. Due
to the coupling, the phases can lock and phase patterns can form.\label{fig:Setup-and-model.}}
\end{figure}

In the system that we study, all oscillators are undergoing motion
on a limit cycle, see Fig. \ref{fig:Setup-and-model.}. We start with
the following equations (``Hopf equations''), which describe the
effective phase and amplitude dynamics of these resonators close to
the limit cycle \cite{Heinrich_Marquardt_Collective_Dynamics,Ludwig_Marquardt_Quantum_Many_Body}:
\begin{align}
\dot{\varphi}_{i}= & -\bar{\Omega}-(A_{i}-\bar{A})\frac{\partial\Omega}{\partial A}(\bar{A})+\frac{F_{i}(t)}{m\Omega(A_{i})A_{i}}\cos(\varphi_{i}),\nonumber \\
\dot{A}_{i}= & -\gamma(A_{i}-\bar{A})+\frac{F_{i}(t)}{m\Omega(A_{i})}\sin(\varphi_{i}).
\end{align}
Here $\bar{\Omega}=\Omega(\bar{A})$ is the frequency at the steady-state
amplitude $\bar{A}$. The frequency-pull due to amplitude changes
has been kept to leading order in the first line, $\gamma$ is the
rate at which the amplitude is forced back to the limit cycle, and
$F_{i}(t)$ is the total force exerted on resonator $i$. For the
physically most relevant spring-like coupling between nearest neighbors,
which will be studied here, the force is given as $ $$F_{i}=k\sum_{\langle j,i\rangle}A_{j}\cos(\varphi_{j})$.
The coupling constant is $k$, and $\langle j,i\rangle$ indicates
nearest neighbor sites.

We consider the case of weak coupling $k/(m\bar{\Omega}^{2})\ll1$
and assume $\gamma/\bar{\Omega}\ll1$, $(\bar{A}/\bar{\Omega})\partial\Omega/\partial A\ll1$.
Then the amplitude fluctuations around the steady state value are
small and the amplitude dynamics can be integrated out (for details
on this step, also about disorder, see \cite{Heinrich_Marquardt_Collective_Dynamics,Ludwig_Marquardt_Quantum_Many_Body}
and the Supplemental Material (SM)). We arrive at effective equations
for the resonator phases. Keeping only the slow phase dynamics (and
assuming identical resonators), we get
\begin{align}
\dot{\varphi}_{i}= & \, C\sum_{\langle j,i\rangle}\cos(\varphi_{j}-\varphi_{i})+S_{1}\sum_{\langle j,i\rangle}\sin(\varphi_{j}-\varphi_{i})\nonumber \\
 & +S_{2}\Big\{\sum_{\langle j,i\rangle}\sum_{\langle k,j\rangle}\big[\sin(2\varphi_{j}-\varphi_{k}-\varphi_{i})-\sin(\varphi_{k}-\varphi_{i})\big]\nonumber \\
 & +\sum_{\langle j,i\rangle}\sum_{\langle k,i\rangle}\sin(\varphi_{k}+\varphi_{j}-2\varphi_{i})\Big\}.\label{eq:OMKM-1}
\end{align}
with $C=k/2m\bar{\Omega}$, $S_{1}=(C\bar{A}/\gamma)(\partial\Omega/\partial A)\vert_{A=\bar{A}}$,
$S_{2}=C^{2}/2\gamma$. We will call Eq.~(\ref{eq:OMKM-1}) the \emph{Hopf-Kuramoto
model}. It has been derived before in the context of optomechanics
\cite{Heinrich_Marquardt_Collective_Dynamics,Ludwig_Marquardt_Quantum_Many_Body},
but it holds generally for a set of weakly coupled Hopf oscillators.
Our aim is to explore the dynamics of this model on a square lattice.

The term $\sin(\varphi_{j}-\varphi_{i})$ of Eq.~(\ref{eq:OMKM-1})
is well known from the Kuramoto model \cite{Kuramoto_original}, or,
equivalently, the XY model \cite{Kosterlitz_Thouless_original}. Here
the term arises from the amplitude-dependence of the frequency \cite{Cross_Lifshitz_Sync_frequency_pulling}.
Both contributions in the first line of Eq.~(\ref{eq:OMKM-1}) have
been derived previously for coupled limit-cycle oscillators, see \cite{Lifshitz_Collective_Dynamics_inBook}.
They are linear in the coupling $k$. In contrast, the prefactor $S_{2}$
is of second order in $k$. However, as discussed below, in realistic
scenarios $\gamma$ and $\partial\Omega/\partial A$ can become small,
such that the regime $S_{2}\sim S_{1},C$ is easily reached. The $S_{2}$-term
can then have a profound influence on the pattern formation dynamics.
The additional contribution also displays next-to-nearest-neighbor
coupling of the phases, in spite of the underlying intrinsic nearest-neighbor
coupling in the lattice assumed here.

We will first set the stage by highlighting several limiting cases
of our model, some of which are known already. For $C=0$, Eq.~(\ref{eq:OMKM-1})
can be rewritten in the form $\dot{\varphi}_{i}=-\partial U/\partial\varphi_{i}$.
Hence, the system will slide down to a minimum of the potential $U$.
In contrast, for non-vanishing $C$, such a potential does not exist
and the system may never reach a stationary state (where $\dot{\varphi}_{i}$
is constant). The limiting case of Eq.~(\ref{eq:OMKM-1}) with $S_{2}=0$
has been studied before \cite{Sakaguchi_Kuramoto_Rotater_Model,Sakaguchi01051988,Kuramoto_Lattice_of_Rings,Kim_Moon_Pattern_Formation,Wiesenfeld_Strogatz_Josephson_arrays}.
This is the Sakaguchi-Kuramoto model, usually written in the form
$\dot{\varphi}_{i}=\, K\sum_{\langle j,i\rangle}\sin(\varphi_{j}-\varphi_{i}+\alpha)$
with $\tan(\alpha)=C/S_{1}$ and $K^{2}=S_{1}^{2}+C^{2}$.$ $

The continuum limit of Eq.~(\ref{eq:OMKM-1}), which is valid for
smooth phase fields, is given by
\begin{equation}
\dot{\varphi}=\mathcal{S}_{1}\Delta\varphi-2\mathcal{S}_{2}\Delta^{2}\varphi-\mathcal{C}(\nabla\varphi)^{2}+4C,\label{eq:OMKM_cont}
\end{equation}
where $\mathcal{S}_{1}=S_{1}a^{2},\ \mathcal{S}_{2}=S_{2}a^{4},\ \mathcal{C}=Ca^{2}$,
with lattice constant $a$. In this model (with $\mathcal{S}_{2}=0$)
it has been found that spirals can develop around singularities in
the phase field \cite{Kuramoto01091976,Kuramoto_Lattice_of_Rings}.
Besides, it has been analyzed in connection with chemical turbulence
in one dimension \cite{Yamada01081976,Kuramoto_Persistent_Propagation_1976,Kuramoto_Instability_and_Turbulence_1980}.

The aim of this paper is to explore pattern formation in the full
model, Eq. (\ref{eq:OMKM-1}), in large two-dimensional arrays. Our
main result is the pattern phase diagram discussed further below.
The patterns we find will determine the phase synchronization dynamics
of limit-cycle oscillators.

\begin{figure}
\includegraphics[width=1\columnwidth]{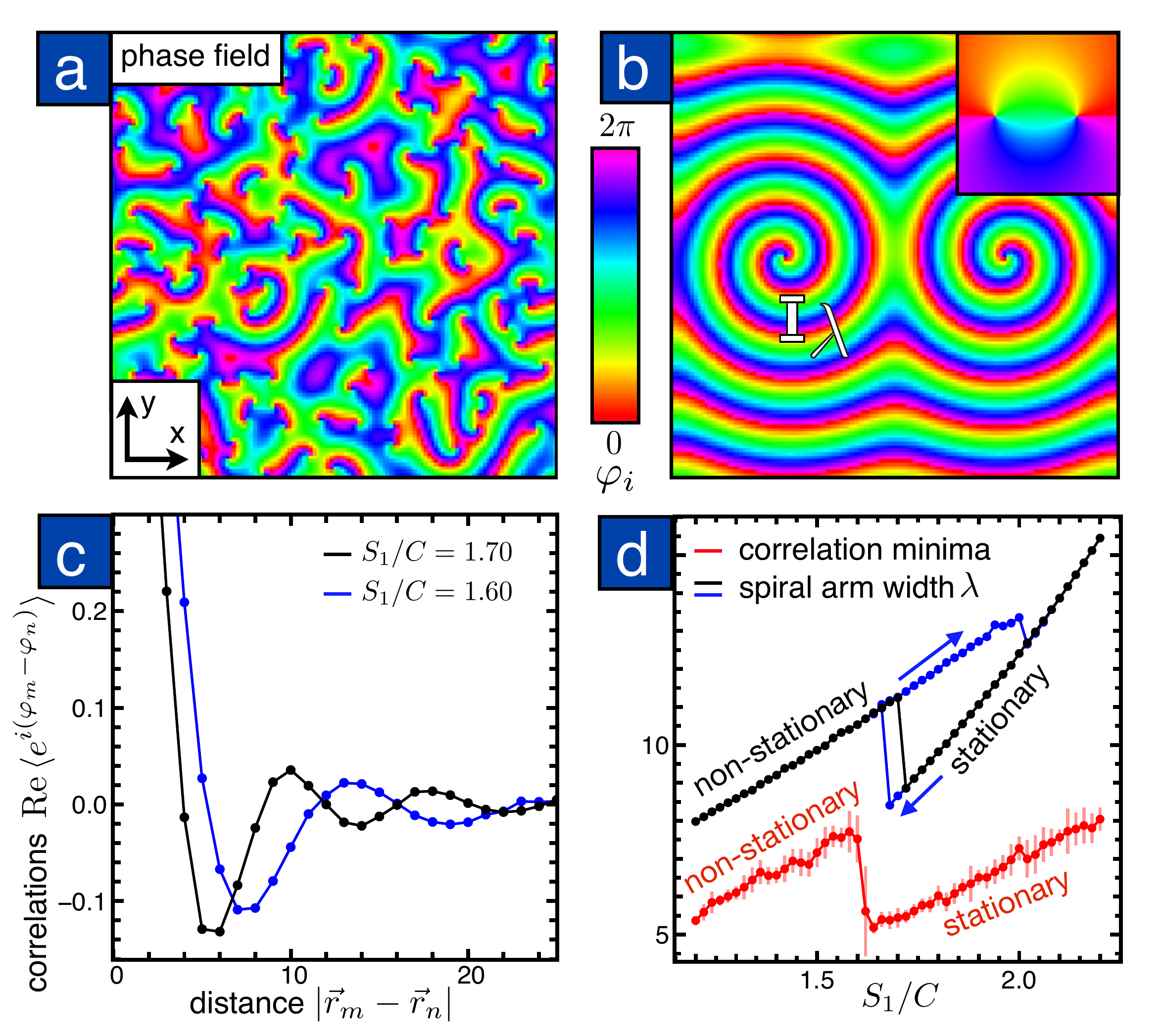}

\caption{Spiral patterns and length scales in the Hopf-Kuramoto model. (a)
Stationary spiral pattern emerging from random initial conditions
for $S_{1}/C=2.0,\ S_{2}/C=0$ on a $N\times N$ square lattice ($N=128$)
with periodic boundary conditions. (b) Vortex-anti-vortex pair (see
inset) winding up to a stationary spiral-anti-spiral pair with a characteristic
spiral arm width $\lambda$. Parameters are like in (a). (c) Spatial
correlations ${\rm Re}\langle\exp(i(\varphi_{m}-\varphi_{n}))\rangle$
as a function of the distance $|\vec{r}_{m}-\vec{r}_{n}|$ (rounded
to the nearest integer). To obtain the data for (d), we extract the
first correlation minimum position from parabolic fits and average
over 10 runs with different random initial conditions. (d) The location
of the first correlation minimum (red) and the spiral arm width $\lambda$
from (b) (black), as a function of the ratio $S_{1}/C$, in units
of the lattice constant. There can be hysteresis (blue).\label{fig:Spiral-patterns-and-length-scale}}
\end{figure}

Our numerical results are mostly obtained from simulations with a
random initial phase field, since that is the natural starting point
in real systems (e.g. after switching on the pump laser or the feedback
driving the oscillators into the limit cycle). After some transient
dynamics, we often find patterns that do not change qualitatively
any more on longer time scales. Moreover, in certain parameter regimes,
we find nontrivial stationary patterns.

\begin{figure*}
\includegraphics[width=1.87\columnwidth]{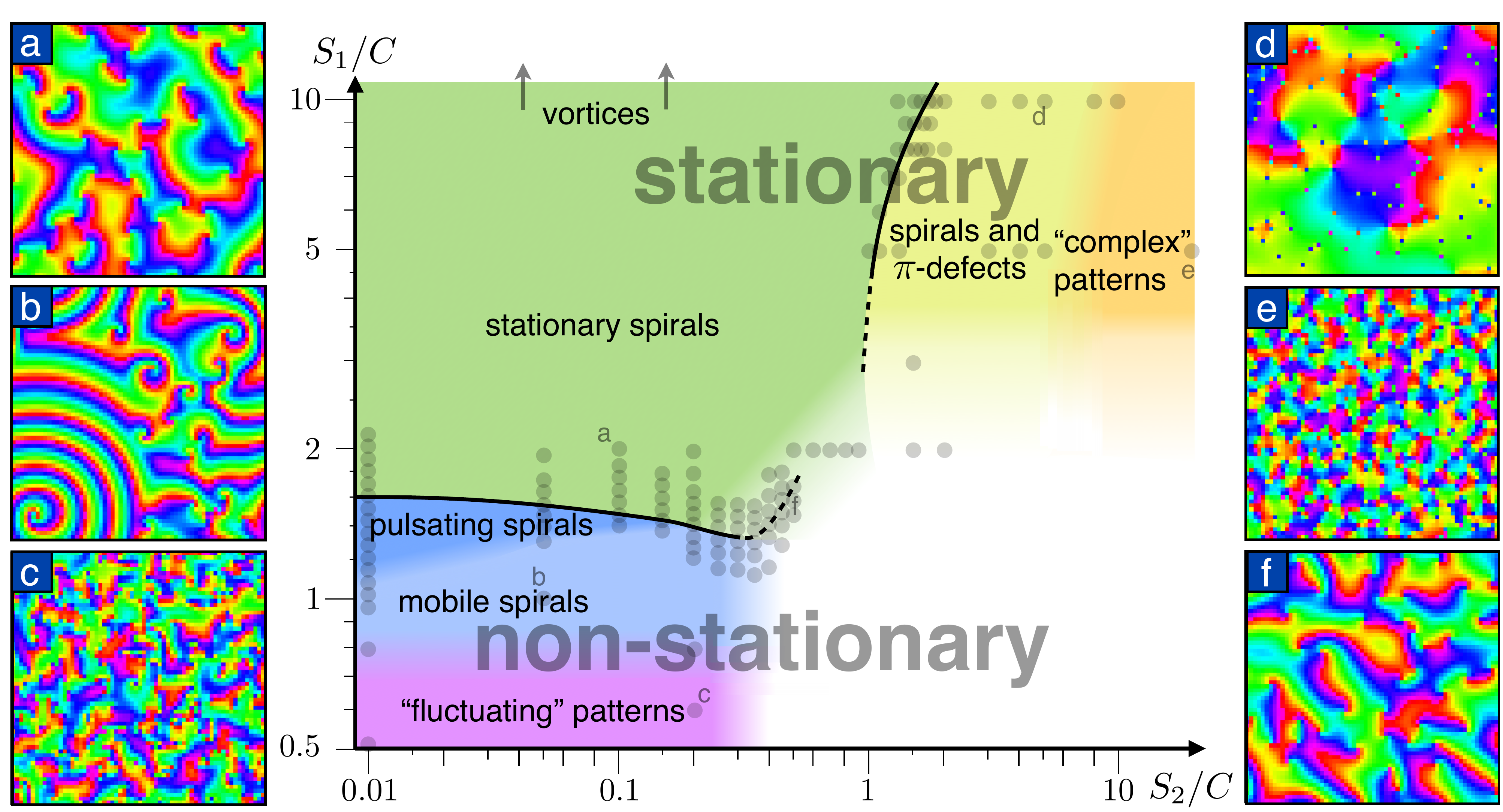}

\caption{Pattern phase diagram of the Hopf-Kuramoto model, Eq. (\ref{eq:OMKM-1}).
Different colors indicate different patterns, which are discussed
in the main text. Sharp transitions occur between stationary spirals
and pulsating/mobile spirals (for small $S_{2}/C$), and in the appearance
of ``$\pi$-defects''. Point markers indicate parameters explored
by numerical simulation. Some typical phase patterns are shown in
the insets (a) to (f).\label{fig:Pattern-phase-diagram}}
\end{figure*}
A typical final pattern of a simulation with large $S_{1}/C$ is shown
in Fig.~\ref{fig:Spiral-patterns-and-length-scale}a. This pattern
is stationary. It consists of many vortex-like ``singularities'',
where the phase changes by $2\pi$ when going around in a closed loop.
These points are surrounded by spiral structures. Spiral patterns
in general are well-known as a recurring motif in pattern formation
\cite{Cross_Pattern_Formation_review,Winfree_Spiral_Waves_BZ}. Since
they form an important part of the patterns we observe, we now briefly
discuss the properties of isolated spirals, produced from an initial
condition with a vortex in the phase field (Fig.~\ref{fig:Spiral-patterns-and-length-scale}b).

It is known that in related models, there is a transition from stationary
spirals to non-stationary spirals, i.e. a situation when the spiral
centers are no longer phase-locked to the bulk of the lattice \cite{Sakaguchi01051988,Paullet_Ermentrout_1994}.
We have discovered that this transition also gives rise to a jump
in the width of the spiral arms, $\lambda$ (Fig.~\ref{fig:Spiral-patterns-and-length-scale}).
Outside of the jump, $\lambda$ increases with increasing $S_{2}/C$
and $S_{1}/C$ (black curve in Fig. \ref{fig:Spiral-patterns-and-length-scale}d).
When sweeping the parameter ratio $S_{1}/C$ up and down, we find
hysteresis in the spiral arm width (blue line in Fig.~\ref{fig:Spiral-patterns-and-length-scale}d).
The precise value at which the jump occurs can then depend on the
parameter sweep rate. Our analysis illustrates that the microscopic
details of the spiral center, on the scale of a few lattice sites,
influence both the spiral arm width and the macroscopic pattern considerably.
Because the structure of the spiral core is complicated, we cannot
provide an analytical prediction for $\lambda$.

We now turn to the statistical properties of the patterns which evolve
out of random initial conditions (see Fig.~\ref{fig:Spiral-patterns-and-length-scale}a),
as they are directly relevant for experiments. The spatial correlations
of the phase field are characterized by the correlator $\langle e^{i(\varphi_{m}-\varphi_{n})}\rangle$,
whose distance dependence is displayed in Fig.~\ref{fig:Spiral-patterns-and-length-scale}c.
We find an oscillatory structure connected to the presence of spiral
arms. On top of that, there is an exponential decay, due to the presence
of many randomly located spiral centers. The position of the first
minimum in the oscillations indicates the distance approximately set
by half a spiral arm width. The dependence of this distance on the
parameter $S_{1}/C$ is shown as the red line in Fig.~\ref{fig:Spiral-patterns-and-length-scale}d.
Again, we find a sudden jump associated with the transition from stationary
to non-stationary spiral centers. We note that the spiral arm width
$\lambda$ determined for isolated spirals does not agree completely
with the length scale extracted from the oscillations of the correlator.
The difference can be traced back to changes in the spiral core produced
by the presence of other nearby spirals.

There are only two dimensionless parameters, $S_{1}/C$ and $S_{2}/C$,
that determine the properties of the final pattern. Therefore, a complete
overview of the various regimes in our model is provided by the ``pattern
phase diagram'' in Fig.~\ref{fig:Pattern-phase-diagram}. This summarizes
the main results of our studies, and we now explain its features.

The transition discussed above, between stationary and non-stationary
spirals, is sharp and can be traced up to intermediate values of $S_{2}/C$.
In addition, we find two classes of non-stationary spirals: ``pulsating''
spirals, where the core keeps orbiting in a small circle around a
fixed location \cite{Paullet_Ermentrout_1994}, and truly mobile spirals
that move through the whole lattice. We will comment on their dynamics
later. We have not observed a sharp transition between the two regimes
(Fig.~\ref{fig:Pattern-phase-diagram}). At larger $S_{2}/C$, the
transition is directly from stationary to mobile spirals.

When decreasing the parameter $S_{1}/C$ even further, we find a crossover
to ``fluctuating'' patterns, see Fig. \ref{fig:Pattern-phase-diagram}c.
These are non-stationary patterns with a complicated phase structure
on the scale of the lattice. For the special case $S_{2}=0$, the
location of the crossover (around $S_{1}/C\sim0.8$) matches the result
obtained in \cite{Kim_Moon_Pattern_Formation}.

The crucial macroscopic length scale of the Hopf-Kuramoto model, i.e.
the spiral arm width, grows with increasing $S_{2}/C$. In view of
that, it is surprising to see microscopic structures appearing at
larger values of this parameter. Indeed, we find a sharp transition
from the domain of ``stationary spirals'' to stationary patterns
that contain ``$\pi$-defects'', see Fig. \ref{fig:Pattern-phase-diagram}d.
These are point defects which are offset by a phase difference of
roughly $\pi$ from the smooth surrounding phase field. The stability
of a single $\pi$-defect on a homogeneous background can be analyzed
by semi-analytical linear stability analysis (in the limit $C\rightarrow0$;
for details, see the SM), which gives the critical value $S_{2}/S_{1}=0.107$.
This defines the asymptote for the transition line in Fig.~\ref{fig:Pattern-phase-diagram}.
Above the critical value, the $\pi$-defect patterns form a fixed
point of the dynamics and can be reached from random initial conditions.
In contrast to the pure spiral patterns, these patterns resolve the
structure of the lattice and hence form a fundamentally different
phase. Obviously, they cannot appear in the continuum model, Eq. (\ref{eq:OMKM_cont}).

When increasing the parameter value $S_{2}/C$ further, the density
of $\pi$-defects increases until we observe a smooth transition to
``complex'' patterns. These are stationary patterns with a complicated
phase structure on the scale of the lattice, see Fig. \ref{fig:Pattern-phase-diagram}e.

Finally, we note that the white region in the phase diagram could
not be accessed due to the significant increase of timescales. Apart
from that, we have discussed all phases in the Hopf-Kuramoto model,
for positive parameters. Changing the sign of $C$ or $S_{1}$ will
not give qualitatively different results: The emerging patterns can
be reconstructed from the patterns discussed above by the transformations
$\varphi_{m,n}\to-\varphi_{m,n}$ for a sign change of $C$, and $\varphi_{m,n}\to-\varphi_{m,n}+(-1)^{m+n}\pi$/2
(``checkerboard gauge'') for a sign change of $S_{1}$. This works
for all values of $S_{2}$. However, changing the sign of $S_{2}$
will lead to different patterns. These involve structure on the scale
of the lattice, where phase differences of roughly $\pi/2$ play an
important role. We will not discuss these patterns, because for coupled
Hopf oscillators $S_{2}$ is positive. 

\begin{figure}
\includegraphics[scale=0.22]{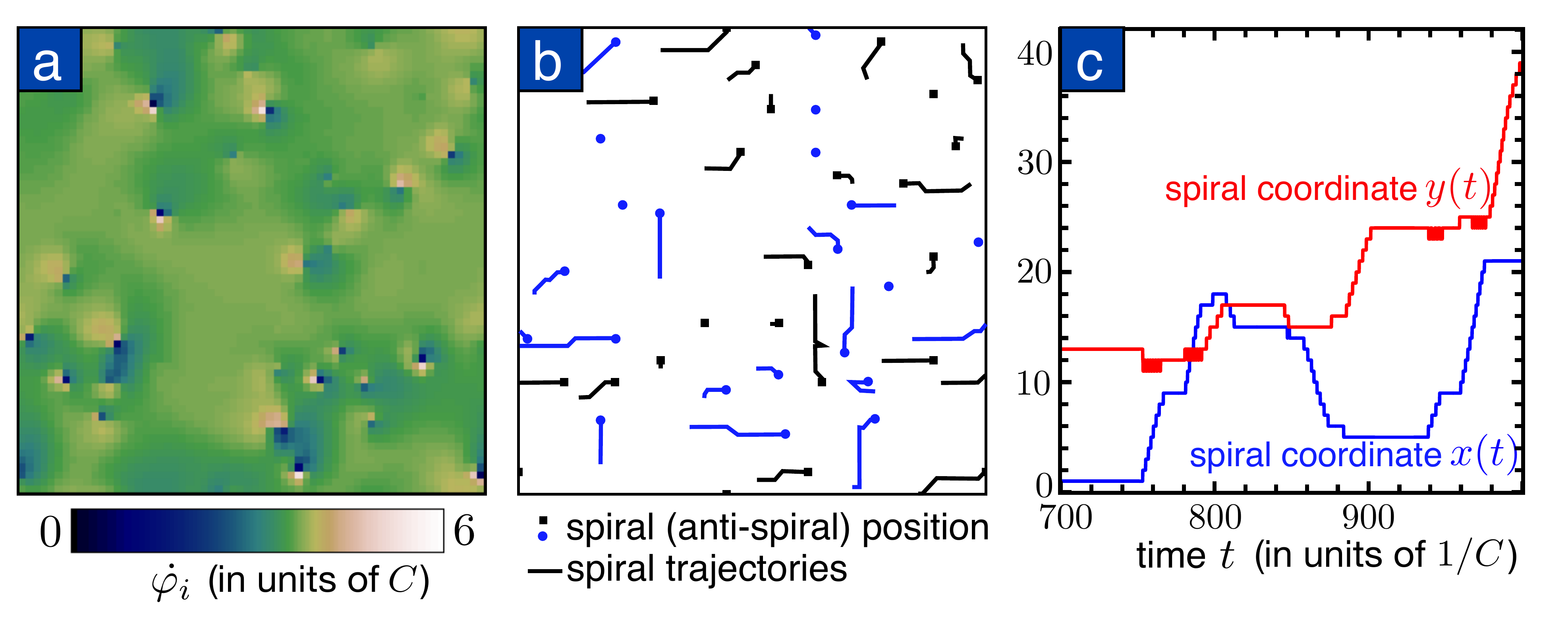}\caption{Spiral motion in the Hopf-Kuramoto model. (a) Instantaneous phase
velocity field $\dot{\varphi}_{i}$ after a (long) integration time
$T=1000/C.$ Mobile spiral centers are visible as local inhomogeneities
in $\dot{\varphi}_{i}$. Some stationary spirals (not visible) exist
in the uniform regions. Parameters are $S_{1}/C=1.6,\ S_{2}/C=0.5,\ N=64$.
The corresponding phase field is shown in Fig.~\ref{fig:Pattern-phase-diagram}f.
(b) Spiral positions at time $T$. The lines are the spiral trajectories
(from $T-15/C$ to $T$; the trajectories have been slightly smoothened
for clarity). (c) Spiral dynamics of a single spiral over a longer
period of time. It can be seen that the spiral remains fixed for some
time before starting to move again (this is usually induced via a
kick by a nearby mobile spiral). The spiral preferably moves in the
cartesian directions set by the lattice.\label{fig:Spiral-movement}}
\end{figure}

We now turn to a more detailed discussion of the spiral motion and
interaction (see also \cite{Aranson_Spiral_Interaction} for the continuum
case, at $S_{2}=0$). Whenever we observe mobile spirals, a fraction
of the spirals and anti-spirals eventually annihilate. In some cases,
they can also be created dynamically. We observe that the spirals
move through the array almost independent of one another for small
$S_{2}/C$, whereas they tend to move in pairs for larger values of
this parameter. For large values of $S_{2}/C$, mobile and stationary
spirals can even coexist, see Fig. \ref{fig:Spiral-movement}. Depending
on initial conditions, the final state can then be non-stationary
or stationary (if all mobile spirals annihilate).

There is also a parameter regime where $\pi$-defects, stationary
and mobile spirals can all be present and interact: Upon the annihilation
of a spiral-anti-spiral pair, a $\pi$-defect can be left behind.
This happens more often for larger values of $S_{2}/C$. When a mobile
spiral approaches a $\pi$-defect, it can induce the dissolution of
the defect into a spiral-anti-spiral pair. However, the mobile spiral
can also move across the defect and make it vanish. All these interactions
play an important role even at late times.

Experimental studies of the patterns discussed in this work could
be implemented by direct local electrical readout of the motion in
electrically coupled nanomechanical resonator arrays \cite{Roukes_Sync_nanomechanical_oscillators},
or by optical readout of the motion in future optomechanical arrays
\cite{Heinrich_Marquardt_Collective_Dynamics,Ludwig_Marquardt_Quantum_Many_Body}
based on optomechanical crystals \cite{Eichenfield_Painter_2009,Safavi-Naeini_Painter_2d_Optomechanical_Crystal_Cavity}
or other platforms. The latter consist of an array of localized mechanical
modes, each of them coupled to one localized optical mode, driven
by a laser. The model parameters could be tuned by varying the laser
power and detuning. Simulations of single optomechanical cells, where
we extracted the phenomenological parameters $\gamma$, $\bar{A}$
and $(\partial\Omega/\partial A)\rvert_{A=\bar{A}}$, suggest that
all the important regions of the pattern phase diagram could be explored.
Near the Hopf bifurcation, $\gamma\lesssim C$ can be reached (since
$\gamma\rightarrow0$), so $S_{2}\gtrsim C$. Furthermore, $S_{2}\gtrsim S_{1}$
holds as well for sufficient coupling $k$, when $\bar{A}(\partial\Omega/\partial A)\vert_{A=\bar{A}}\lesssim C$.
The motion can be read out by observing the light scattered from the
sample. The intensity of the light scattered with wave vector transfer
$\vec{q}$ is related to the structure factor (see SM), i.e. the spatial
Fourier transform (at $\vec{q}$) of the phase correlator $\left\langle e^{i(\varphi_{l}-\varphi_{j})}\right\rangle _{t}$.
In a typical experiment, the frequencies will be disordered, but first
simulations (with standard deviation $0.1C$) do not show qualitative
changes of the patterns we discussed. However, initially mobile spirals
could be pinned at sites with lower frequencies \cite{Sakaguchi01051988}. 

The variety of patterns summarized in Fig.~\ref{fig:Pattern-phase-diagram}
are important for synchronization dynamics and applications. For example,
finite phase-differences across the array (in stationary patterns)
will reduce the total power output of the collective oscillator, and
the mere presence of spirals can reduce the robustness against noise
\cite{Allen_Cross_Frequency_precision_with_spirals}. Finite frequency-differences
(in non-stationary patterns) reduce the frequency stability. Tuning
the parameters into suitable regions will optimize the array's properties.
Future theoretical studies of the Hopf-Kuramoto model could include
noise, which may lead to interesting effects, as discussed for similar
models in \cite{Acebron_Spigler_Review}. In that context, as well
as in the deterministic case, the role of spiral motion and interaction
could be analyzed in more detail.
\begin{acknowledgments}
We acknowledge support from an ERC Starting Grant, the DARPA program
ORCHID, and the ITN cQOM. We thank Ron Lifshitz for helpful discussions.
\end{acknowledgments}

\clearpage

\global\long\def\theequation{S.\arabic{equation}}

\global\long\def\thefigure{S.\arabic{figure}}

\setcounter{equation}{0}

\setcounter{figure}{0}

\thispagestyle{empty}
\onecolumngrid
\begin{center}
{\fontsize{12}{12}\selectfont
\textbf{\textit{Supplemental Material} for ``Pattern phase diagram for 2D
arrays of coupled limit-cycle oscillators''\\
[5mm]}}
{\normalsize Roland Lauter,$^{1,2,*}$ Christian Brendel,$^1$ Steven J.~M.~Habraken,$^1$ and  Florian Marquardt$^{1,2}$\\[1mm]} 
{\fontsize{9}{9}\selectfont  
\textit{$^1$Institut f\"ur Theoretische Physik II, Friedrich-Alexander-Universit\"at \\
Erlangen-N\"urnberg, Staudtstr. 7, 91058 Erlangen, Germany\\
$^2$Max Planck Institute for the Science of Light, G\"unther-Scharowsky-Stra{\ss}e 1/Bau 24, 91058 Erlangen, Germany }} \vspace*{6mm}  \end{center}   \normalsize

\section{Derivation of the Hopf-Kuramoto model}

\noindent In this section, we derive the Hopf-Kuramoto model from
the following general Hopf equations
\begin{align}
\dot{\varphi}_{i} & =-\bar{\Omega}_{i}-\frac{\partial\Omega_{i}}{\partial A_{i}}\rvert_{A_{i}=\bar{A}}(A_{i}-\bar{A}_{i})+\frac{F_{i}(t)}{m_{i}\Omega_{i}(A_{i})A_{i}}\cos\varphi_{i},\label{phaseeq}\\
\dot{A}_{i} & =-\gamma(A_{i}-\bar{A}_{i})+\frac{F_{i}(t)}{m_{i}\Omega_{i}(A_{i})}\sin\varphi_{i}.\label{ampeq}
\end{align}
Here, $\Omega_{i}(A_{i})$ is the amplitude-dependent frequency of
the oscillator at site $i$, $m_{i}$ is its mass, $\bar{A}_{i}$
is its steady-state amplitude and $\bar{\Omega}_{i}=\Omega_{i}(\bar{A}_{i})$
is the frequency at the steady-state amplitude. Other symbols have
the same meaning as in the main text. The second term on the right-hand
side of Eq. (\ref{phaseeq}) arises from the expansion of $\Omega_{i}(A_{i})$
around the steady-state amplitude $\bar{A}_{i}$. For reasons that
will become clear later, $\Omega_{i}(A_{i})$ has not been expanded
in the force terms in Eqs. (\ref{phaseeq}) and (\ref{ampeq}). We
assume that the oscillators are coupled by spring-like nearest-neighbor
couplings so that the forces $F_{i}(t)$ are given by
\begin{equation}
F_{i}=\sum_{\langle j,i\rangle}k_{ij}x_{j}=\sum_{\langle j,i\rangle}k_{ij}A_{j}\cos\varphi_{j},
\end{equation}
where $\langle j,i\rangle$ denotes the nearest neighbors $j$ of
site $i$ and $k_{ij}=k_{ji}$ are spring constants.

The derivation of the Hopf-Kuramoto model involves the adiabatic elimination
of the amplitude fluctuations $\delta A_{i}=A_{i}-\bar{A}_{i}$, as
well as leading-order expansions in the dimensionless, small parameters
$k_{ij}/(m_{i}\bar{\Omega}_{i}^{2})$, $(\bar{A}_{i}/\Omega_{i})\partial\Omega_{i}/\partial A_{i}$
and $\gamma/\bar{\Omega}_{i}$. These parameters and the relative
amplitude fluctuations $\delta A_{i}/\bar{A}$ are assumed to be of
the same order of smallness. Below, we will also only keep slowly
varying terms. The derivation can also be found in \cite{Heinrich_Marquardt_Collective_Dynamics2}.
For some more details, see the Supplemental Material of \cite{Ludwig_Marquardt_Quantum_Many_Body2}.

In order to eliminate the amplitude fluctuations, we rewrite Eq. (\ref{ampeq})
in terms of the amplitude fluctuations and formally integrate the
equation
\begin{equation}
\delta\dot{A}_{i}=-\gamma\;\delta A_{i}+\frac{\sin\varphi_{i}}{m_{i}\Omega_{i}(A_{i})}\sum_{\langle j,i\rangle}k_{ij}\left(\bar{A}_{i}+\delta A_{i}\right)\cos\varphi_{j},
\end{equation}
to obtain the long-time limit result
\begin{equation}
\delta A_{i}(t)=\frac{1}{m_{i}\Omega_{i}(A_{i})}\int_{-\infty}^{t}dt'\; e^{-\gamma(t-t')}\sin\varphi_{i}(t')\sum_{\langle j,i\rangle}k_{ij}\left(\bar{A}_{i}+\delta A_{i}(t')\right)\cos\varphi_{j}(t').
\end{equation}
Since the integrand is proportional to $k_{ij}$, to leading order
it suffices to evaluate $\varphi_{i}(t')$ to zeroth order in the
expansion parameters, i.e. $\varphi_{i}(t')\simeq\varphi_{i}(t)-\Omega_{i}(t-t')$.
Thus, we find
\begin{equation}
\delta A_{i}\simeq\sum_{\langle j,i\rangle}\frac{\bar{A}_{i}k_{ij}}{m_{i}\Omega_{i}(A_{i})}\int_{-\infty}^{t}dt'\; e^{-\gamma(t-t')}\sin\Big(\varphi_{i}(t)-\Omega_{i}(t-t')\Big)\cos\Big(\varphi_{j}(t)-\Omega_{j}(t-t')\Big).
\end{equation}
The integral can easily be evaluated. To leading order in $\gamma/\bar{\Omega}_{i}$,
the result reduces to
\begin{equation}
\delta A_{i}\simeq\sum_{\langle j,i\rangle}\frac{\bar{A}_{i}k_{ij}}{m_{i}\Omega_{i}(A_{i})}\frac{\sin(\varphi_{i}-\varphi_{j})}{2\gamma}.\label{adel}
\end{equation}
To first order in the amplitude fluctuations, the equations of motion
for the oscillator phases (\ref{phaseeq}) can be expanded as
\begin{equation}
\dot{\varphi}_{i}\simeq-\bar{\Omega}_{i}-\frac{\partial\Omega_{i}}{\partial A_{i}}\rvert_{A_{i}=\bar{A}_{i}}\delta A_{i}+\frac{\cos\varphi_{i}}{m_{i}\Omega_{i}(A_{i})}\sum_{\langle j,i\rangle}k_{ij}\left(1+\frac{\delta A_{j}-\delta A_{i}}{\bar{A}}\right)\cos\varphi_{j}.
\end{equation}
Corrections to $\Omega_{i}(A_{i})\simeq\Omega_{i}(\bar{A}_{i})=\bar{\Omega}_{i}$
are proportional to both $(\bar{A}_{i}/\Omega_{i})\partial\Omega_{i}/\partial A_{i}$
and $\delta A_{i}/\bar{A}_{i}$ so that they are of second order.
In the second term on the right-hand side, these are significant,
but, since in the third term they are multiplied by another $k_{ij}$,
they can be neglected there. Inserting (\ref{adel}), again replacing
$\Omega_{i}(A_{i})$ in the denominator by $\bar{\Omega}_{i}$, we
finally obtain
\begin{align}
\dot{\varphi}_{i}\simeq & -\bar{\Omega}_{i}+\frac{\partial\Omega_{i}}{\partial A_{i}}\rvert_{A_{i}=\bar{A}_{i}}\sum_{\langle j,i\rangle}\frac{\bar{A}_{i}k_{ij}}{m_{i}\bar{\Omega}_{i}}\frac{\sin(\varphi_{j}-\varphi_{i})}{2\gamma}+\sum_{\langle j,i\rangle}\frac{k_{ij}\cos(\varphi_{j}-\varphi_{i})}{2m_{i}\bar{\Omega}_{i}}\\
 & +\sum_{\langle j,i\rangle}\sum_{\langle k,j\rangle}\frac{k_{ij}^{2}}{8\gamma m_{i}^{2}\bar{\Omega}_{i}^{2}}\bigg(\sin(2\varphi_{j}-\varphi_{i}-\varphi_{k})-\sin(\varphi_{k}-\varphi_{i})\bigg)+\sum_{\langle j,i\rangle}\sum_{\langle k,i\rangle}\frac{k_{ij}^{2}}{8\gamma m_{i}^{2}\bar{\Omega}_{i}^{2}}\bigg(\sin(\varphi_{k}+\varphi_{j}-2\varphi_{i})\bigg),\nonumber 
\end{align}
where we have also only kept slowly varying contributions by applying
the approximations $\cos\varphi_{i}\cos\varphi_{j}\simeq\frac{1}{2}\cos(\varphi_{i}-\varphi_{j})$,
$\cos\varphi_{i}\cos\varphi_{j}\sin(\varphi_{j}-\varphi_{k})\simeq\frac{1}{4}\{\sin(\varphi_{i}-\varphi_{k})-\sin(\varphi_{i}+\varphi_{k}-2\varphi_{j})\}$
and $\cos\varphi_{i}\cos\varphi_{j}\sin(\varphi_{i}-\varphi_{k})\simeq\frac{1}{4}\{\sin(2\varphi_{i}-\varphi_{j}-2\varphi_{k})-\sin(\varphi_{k}-\varphi_{j})\}$.
In the special case of identical oscillators $\bar{\Omega}_{i}=\bar{\Omega}$,
$m_{i}=m$ and $\bar{A}_{i}=\bar{A}$ for all $i$ and uniform couplings
$k_{ij}=k$ for all neighbors $i,j$, this obviously reduces to equation
(2) in the main text, where we have also neglected the trivial term
$-\bar{\Omega}$ on the right hand side.

\section{Semi-analytical stability analysis of point defects}

In this section, we present the analysis of the stability of a single
$\pi$-defect on a homogeneous background phase field for the Hopf-Kuramoto
model (Eq. (2) in the main text) with $C=0$. For this case, the aforementioned
phase configuration, which we call $\varphi^{0}$, is a fixed point
of the dynamics, i.e. $\dot{\varphi}_{i}^{0}=0$ for all sites $i$.
Besides, the equation of motion can be written as $\dot{\varphi}_{i}=-\frac{\partial}{\partial\varphi_{i}}U$
with the potential
\[
U(\varphi_{1},...,\varphi_{N^{2}})=\ \sum_{i}\Big\{\sum_{\langle j,i\rangle}\frac{S_{1}}{2}(1-\cos(\varphi_{j}-\varphi_{i}))+S_{2}\Big[\sum_{\langle j,i\rangle}\sin(\varphi_{j}-\varphi_{i})\Big]^{2}\Big\}.
\]

\begin{figure}
\includegraphics[scale=0.4]{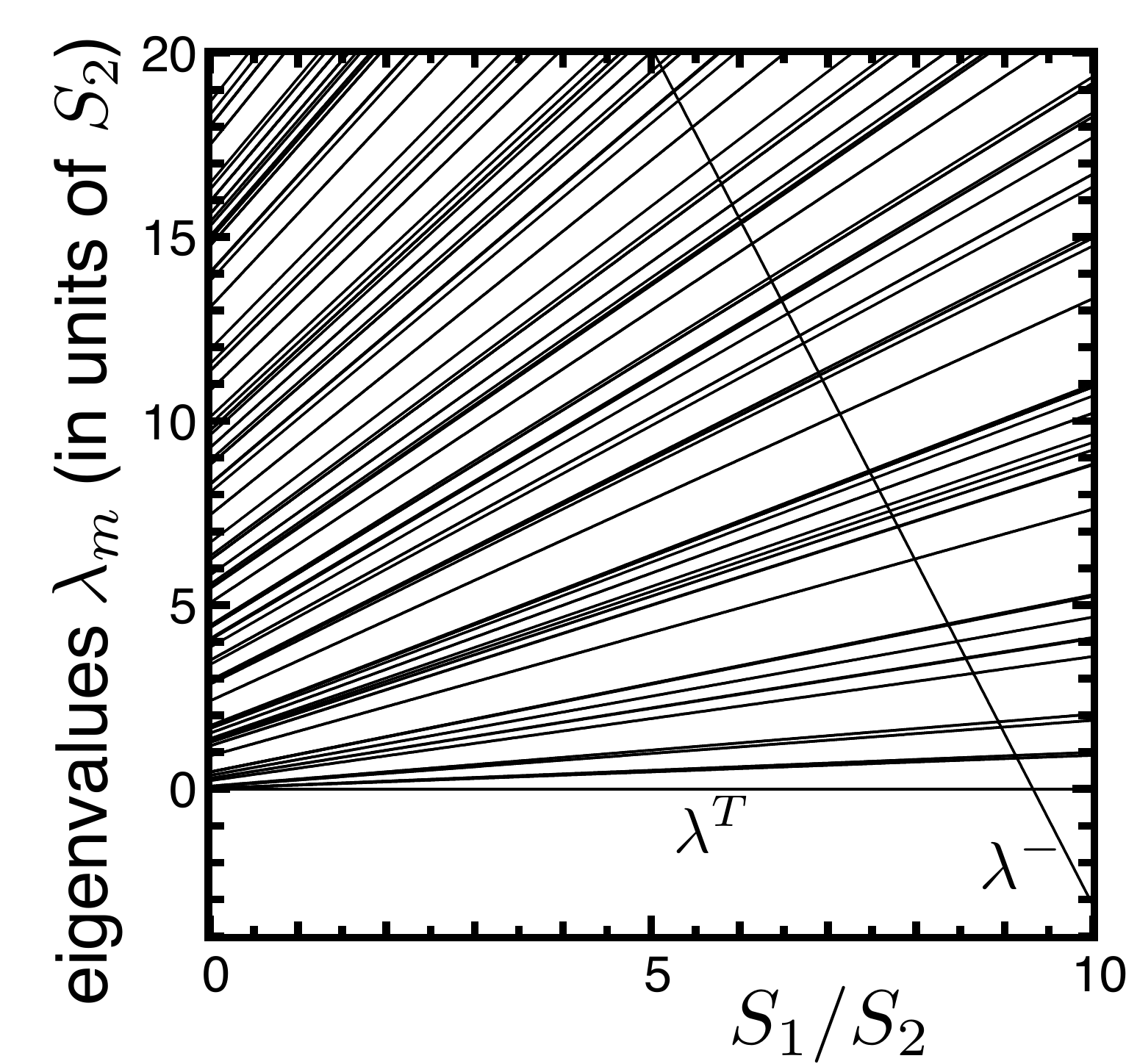}

\caption{The eigenvalues $\lambda_{m}$ of the Hessian $\partial^{2}U/(\partial\varphi_{i}\partial\varphi_{j})$
for a homogeneous phase configuration with a single $\pi$-defect,
as a function of the parameter $S_{1}/S_{2}$ for a lattice of size
$20$x$20$. Note the eigenvalue $\lambda^{-}$ leading to the instability
of the $\pi$-defect. The zero eigenvalue $\lambda^{T}$ belongs to
the translational mode of the phase configuration.\label{fig:stability_analysis}}
\end{figure}

We calculate the Hessian $\partial^{2}U/(\partial\varphi_{i}\partial\varphi_{j})$
and evaluate its eigenvalues for the phase configuration $\varphi^{0}$
numerically. If at least one of the eigenvalues is negative, the configuration
is unstable. A single eigenvalue, corresponding to the translational
mode of the system, might vanish without disturbing our analysis.
We always find this zero eigenvalue. For small values of $S_{1}/S_{2}$,
all the other eigenvalues are positive, which means that $\pi$-defects
are stable (see Fig. \ref{fig:stability_analysis}). With increasing
$S_{1}/S_{2}$, the eigenvalues change linearly with this parameter.
A single eigenvalue $\lambda^{-}(S_{1}/S_{2})$ has a negative slope,
so it becomes negative at some critical value $(S_{1}/S_{2})^{c}\approx9.34$,
rendering the phase configuration unstable. This gives the (inverse)
value $S_{2}/S_{1}\approx0.107$ given in the main text.

\begin{figure}
\includegraphics[width=0.9\columnwidth]{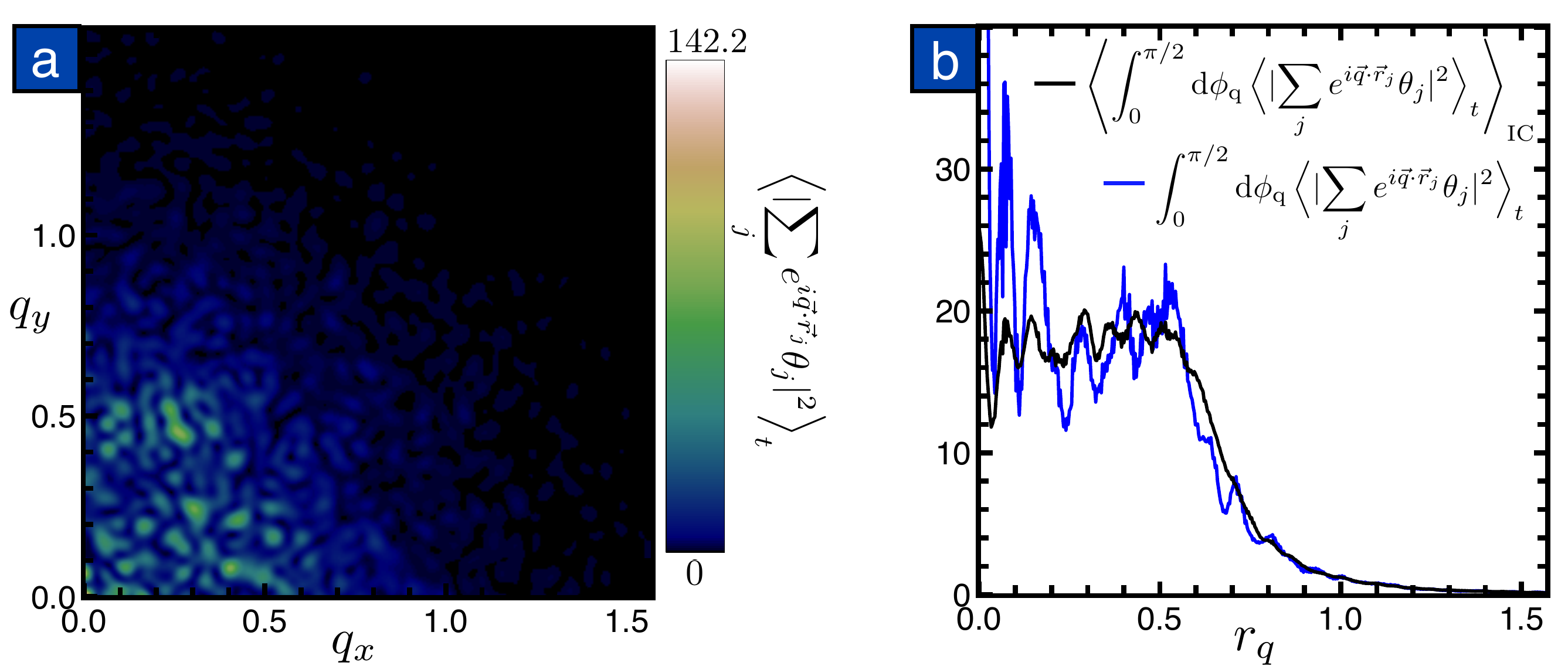}

\caption{(a) Partial intensity $\Big\langle\lvert\sum_{j}e^{i\vec{q}\cdot\vec{r}_{j}}\theta_{j}\rvert^{2}\Big\rangle_{t}$
of the light reflected from the phase field given in Fig. 2a in the
main text. Parameter $\theta^{\mathrm{max}}=0.01$. (b) Partial intensity
$\Big\langle\lvert\sum_{j}e^{i\vec{q}\cdot\vec{r}_{j}}\theta_{j}\rvert^{2}\Big\rangle_{t}$
in dependance on the radial coordinate $q_{r}=\sqrt{q_{x}^{2}+q_{y}^{2}}$,
averaged over 11 different random initial phase configurations (black)
and for a single phase configuration as in (a) (blue).\label{fig:intensity_and_FT}}
\end{figure}

\section{Read-out of the mechanical resonator phase field}

Here, we will show how the intensity of the light reflected from an
optomechanical array is related to the spatial Fourier transform of
the phase correlator. The intensity of the light reflected from an
optomechanical array with lattice sites at $\vec{r}_{j}$ is given
as
\begin{equation}
\lvert E(\vec{q})\rvert^{2}/\lvert E^{\mathrm{in}}\rvert^{2}=\lvert\sum_{j}e^{-i\vec{q}\cdot\vec{r}_{j}}e^{i\theta_{j}}\rvert^{2}.\label{eq:intensity_total}
\end{equation}

The phase of the light reflected from site $j$ is $\theta_{j}=\theta^{\mathrm{max}}\cos(\varphi_{j})$,
where $\theta^{\mathrm{max}}$ depends on the system parameters$ $.
If the mechanical frequency is much smaller than the cavity intensity
decay rate $\kappa$, then $\theta^{\mathrm{max}}=AG/\kappa$, with
the mechanical amplitude $A$ and the optical frequency shift per
displacement $G$ \cite{Aspelmeyer_Marquardt_Review}. For small $\theta^{\mathrm{max}}$,
Eq. (\ref{eq:intensity_total}) can be expanded and we get
\begin{equation}
\lvert E(\vec{q})\rvert^{2}/\lvert E^{\mathrm{in}}\rvert^{2}\approx\sum_{j,l}e^{-i\vec{q}\cdot(\vec{r}_{l}-\vec{r}_{j})}(1+i(\theta_{l}-\theta_{j})-\frac{1}{2}(\theta_{l}-\theta_{j})^{2}).
\end{equation}

We average over time, use $\langle\theta_{j}\rangle_{t}=0$ and $\langle\theta_{j}^{2}\rangle_{t}=(\theta^{\mathrm{max}})^{2}/2$
, and arrive at
\begin{equation}
\Big\langle\lvert E(\vec{q})\rvert^{2}/\lvert E^{\mathrm{in}}\rvert^{2}\Big\rangle_{t}=(1-\frac{(\theta^{\mathrm{max}})^{2}}{2})\lvert\sum_{j}e^{i\vec{q}\cdot\vec{r}_{j}}\rvert^{2}+\Big\langle\lvert\sum_{j}e^{i\vec{q}\cdot\vec{r}_{j}}\theta_{j}\rvert^{2}\Big\rangle_{t}.
\end{equation}

For large arrays, the first term will only give contributions very
close to $\vec{q}=0$. For small arrays, these contributions may have
to be eliminated by calibrating the measurement device with a known
phase field. The second term can be evaluated to give
\begin{equation}
\Big\langle\lvert\sum_{j}e^{i\vec{q}\cdot\vec{r}_{j}}\theta_{j}\rvert^{2}\Big\rangle_{t}=\frac{(\theta^{\mathrm{max}})^{2}}{2}\,\sum_{j,l}e^{-i\vec{q}\cdot(\vec{r}_{l}-\vec{r}_{j})}\mathrm{Re}\langle e^{i(\varphi_{l}-\varphi_{j})}\rangle_{t}.\label{eq:intensity_and_FT}
\end{equation}

On the right-hand side of this equation, the discrete Fourier transform
of the correlations in the system appears. We have analyzed similar
correlation functions in connection with the spiral length scale,
see Fig. 2 in the main text. From Eq. (\ref{eq:intensity_and_FT})
we see that we can learn about the correlations by detecting the intensity
of the reflected light. An example for the part of the detected light
intensity that is given in Eq. (\ref{eq:intensity_and_FT}) is given
in Fig. \ref{fig:intensity_and_FT}.

\end{document}